\documentclass[twocolumn]{aastex6}
\usepackage{graphics,epsf}
\usepackage{amsmath}                
\usepackage{amsfonts}               
\usepackage{amssymb}                
\usepackage{epsfig}                 
\usepackage{graphicx}
\def \kms{\rm{km~s^{-1}}}

\def \s{~\rm{s}}
\def \km{~\rm{km}}

\def \K{~\rm{K}}

\begin{document}

\title{Orbital Parameters for the $250 M_\odot$ Eta Carinae Binary System}
\author{Amit Kashi}
\affil{Minnesota Institute for Astrophysics, University of Minnesota, 116 Church St. SE. Minneapolis, MN 55455, USA}
\affil{Department of Physics, Technion -- Israel Institute of Technology, Haifa 32000, Israel}
\email{kashi@astro.umn.edu}
\author{Noam Soker}
\affil{Department of Physics, Technion -- Israel Institute of Technology, Haifa 32000, Israel}
\email{soker@physics.technion.ac.il}

\shorttitle{The $250 M_\odot$ Binary System $\eta$ Car} 
\shortauthors{A. Kashi \& N. Soker}

\begin{abstract}
We show that recent observations of \ion{He}{1} and \ion{N}{2}
lines of $\eta$ Carinae may provide support for an orbital orientation where the
secondary star is closest to us at periastron passages. This
conclusion is valid both for the commonly assumed masses of the
two stars, and for the higher stellar masses model where the
very massive evolved primary star mass is $M_1=170
M_\odot$ and its hot secondary star mass is $M_2 = 80 M_\odot$.
The later model better explains the change in the orbital period
assuming that the ninetieth century Great Eruption was powered by
accretion onto the secondary star. Adopting the commonly used high
eccentricity $e \simeq 0.9$ and inclination $i=41^\circ$, we
obtain a good fit to newly released Doppler shift observations of
\ion{He}{1} emission and absorption lines assuming they are
emitted and absorbed in the acceleration zone of the secondary
stellar wind. Our model in which the secondary star is in the
foreground at periastron is opposite to the view presented
recently in the literature.
\end{abstract}


\section{INTRODUCTION}
\label{sec:intro}

$\eta$ Car is a binary system \citep{Damineli1996,
Dminelietal1997} composed  of a very massive primary star
\citep{DavidsonHumphreys1997} and a
hotter and less luminous evolved main sequence secondary star.
Despite two decades of detailed observations (e.g.,
\citealt{Smithetal2000, DuncanWhite2003, Whitelocketal2004,
Corcoran2005, Davidsonetal2005, Davidsonetal2015, Smith2006,
Hamaguchietal2007, Daminelietal2008b, Corcoranetal2010,
Martinetal2010, Mehneretal2010, Mehneretal2012, Mehneretal2015,
Abrahametal2014, Hamaguchietal2014a, Hamaguchietal2014b} ) and
modelling (e.g., \citealt{Pittardetal1998, Soker2001,
PittardCorcoran2002, Akashietal2006, Akashietal2013,
KashiSoker2009a, Okazakietal2008, Smith2010, Grohetal2012,
Maduraetal2013, Maduraetal2015, Clementeletal2015b}) there are
disagreements over two important properties of the binary system.
These are the masses of the two stars and the orientation of the
eccentric orbit.

The low-masses model assumes that $\eta$ Car is at its
Eddington luminosity limit and a mass of $>120 M_\odot$ is derived
\citep{Hillieretal2001}. Some studies take it to be the mass of
the primary star, while other assumes that the combined masses of
the two stars amounts to that mass with $M_1 \geq 90 M_\odot$ and
$M_2 \geq 30 M_\odot$ (e.g., \citealt{Okazakietal2008, Clementeletal2015b}).

The high-masses model was developed by us \citep{KashiSoker2010a}
under the assumption that most of the extra energy released during
the 1837--1856 Great Eruption (GE) of $\eta$ Car originated from high
accretion rate onto the secondary star.
There are models that explain the GE with only one star (e.g., \citealt{MattBalick2004}), and others that use
three stars (\citealt{LivioPringle1990}; \citealt{PortegiesZwartvandenHeuvel2016}).

The GE was agiant eruption and a SN Impostor, which is the group of
eruptive massive stars within the more general groups of
intermediate luminosity optical transients (ILOTs;
\citealt{KashiSoker2015}). In \cite{KashiSoker2015} we presented
the high-accretion-powered ILOTs (HAPI) model, according to which
all ILOTs are powered by high accretion rate onto a main sequence
(MS), or a star slightly evolved off the MS \citep{KashiSoker2010b}.
According to the HAPI model, then, the
luminosity peaks of the Great Eruption resulted from accretion
onto the companion close to periastron passages. Those peaks are
$\approx 5.1$--$5.2$ years apart. If one assumes that
the peaks are related to peristron passage, and occur about the same
time after (or before) periastron, the separation between the peaks
can be inferred as the orbital period.
The second peak came after the eruption has started, and mass was lost.
Therefore the orbital period is smaller than the time interval between the peaks.
We adopted a value of $5.1$ years. The present orbital period is
$5.54$ years, suggesting that during the Great Eruption
the orbital period has increased. The orbital period
changed as the secondary accreted mass, and as mass was lost from
the binary system, both as a wind from the primary star and as
jets from the secondary star.

In that study \citep{KashiSoker2010a} we found that for the HAPI
mechanism to work within the orbital changes constraints, the two
stars should have a significantly larger masses than $120
M_\odot$. The masses should be in the range of $M_1 \simeq 150$--$200 M_\odot$
and $M_2 \simeq 60$--$90 M_\odot$. This high-masses
model is further supported by evolution of massive stars on the HR
diagram. The calculations of \cite{Figeretal1998} show that a
zero-age main sequence (ZAMS) star with an initial mass of $M_{\rm
ZAMS} \simeq 230 M_\odot$ is required to explain the present
luminosity of the primary of $\eta$ Car.
Indeed, modern stellar evolution tracks of $120 M_\odot$ stars even when rotation is considered
(\citealt{Ekstrometal2012}; \citealt{Georgyetal2012}) do not reach that luminosity,
and more massive models of $M_{\rm ZAMS} \simeq 250 M_\odot$ are required, as found by \cite{Chenetal2015}.
The primary lost large amounts of mass to the bipolar nebula around $\eta$ Car, the
Homunculus (\citealt{SmithFerland2007, Gomezetal2010}). When the large
mass loss along the evolution is considered, a present mass of
$\sim 150-200 M_\odot$ is compatible with the ZAMS mass inferred above.

In the debate on the orientation of the binary system, the two
sides are holding a literally $\sim 180^\circ$ opposite views.
One side holds that during periastron passages the primary star is
closer to us ($\omega \simeq 240^\circ$--$270^\circ$; e.g.,
\citealt{Ipingetal2005, Nielsenetal2007, Daminelietal2008b,
Henleyetal2008, Parkinetal2009, Grohetal2010, Gulletal2011,
Maduraetal2012, Clementeletal2015a, Richardsonetal2015, Teodoroetal2016}, while
the other side holds the view that during periastron passages the
secondary star is at its closest location to us ($\omega \simeq
90^\circ$; e.g., \citealt{Abrahametal2005, Falcetaetal2005,
AbrahamFalceta2007, KashiSoker2008, KashiSoker2009b,
KashiSoker2011, Tsebrenkoetal2013}).

The disagreement on the orientation stems mainly from the controversy on the
source of some emission and absorption lines, in particular the
\ion{He}{1} lines. In a recent paper \cite{Richardsonetal2015} present
a study of the variations of some spectral lines with the phase of
the binary orbit, based on observations with the CTIO 1.5m telescope.
They attributed the \ion{He}{1} lines to the primary star.
In \cite{KashiSoker2007}, on the other hand, we assumed that
the \ion{He}{1} lines observed by \cite{Nielsenetal2007} are formed in
the acceleration zone of the wind blown by the secondary star. We
calculated the Doppler shift variations of the lines as a function
of orbital phase with the low-masses model of $\eta$ Car, and
found that a good fit is obtained if the \ion{He}{1} lines are formed in
the region where the secondary wind speed is $v_{\rm zone} = 430
\km \s^{-1}$.

In the present study we combine the high-masses model of $\eta$
Car with the assumption that the \ion{He}{1} lines originate in the
acceleration zone of the secondary stellar wind, and try to fit
the new Doppler shifts presented by \cite{Richardsonetal2015}.

\section{THE CONTROVERSY OF THE HELIUM I LINES}
 \label{sec:model}

The origin of the visible \ion{He}{1} P~Cyg lines
($\lambda7065\rm{\AA}$, $\lambda5876\rm{\AA}$,
$\lambda5015\rm{\AA}$, $\lambda4992\rm{\AA}$, and
$\lambda4471\rm{\AA}$) in the binary system $\eta$ Car is in
dispute, with different researchers attributing it to different
regions, e.g., the primary star (e.g.,
\citealt{Falcetaetal2007, Humphreysetal2008, Richardsonetal2015}).
We attribute the \ion{He}{1} P~Cyg lines to the acceleration zone
of the secondary's wind \citep{KashiSoker2007, KashiSoker2008}. A
similar dispute exists for the
\ion{N}{2}~$\lambda\lambda5668$--$5712\rm{\AA}$ line
\citep{Mehneretal2011a}. While \cite{Mehneretal2011a} argued that
it cannot come from the secondary star, we noted that the
\ion{N}{2} lines closely follow the behavior of the \ion{He}{1}
lines, and attributed it to the secondary wind \citep{KashiSoker2011}.

The effective temperature of the secondary was estimated be $T_{\rm{eff},2} \sim
34\,000$--$38\,000 \K$ by \cite{Verneretal2005}, and more recently
$T_{\rm{eff},2} \sim 40\,000$ by \cite{Mehneretal2010}.
The main arguments for the \ion{He}{1} P~Cyg lines origin in the
acceleration zone of the wind blown by the secondary star are as
follows \citep{KashiSoker2011}.
\begin{enumerate}
 \item The Doppler shift of the P~Cyg absorption components follows
very well the secondary's orbit, as we show in section
\ref{sec:orbital} for the high-masses model, and as was shown
before for the low-masses model \citep{KashiSoker2007,
KashiSoker2008}.
 \item The lines are known to originate in stars with temperatures well above $30\,000 \K$, mostly in hydrogen deficient stars (e.g., \citealt{Leuenhagenetal1996}; \citealt{CrowtherBohannan1997}; \citealt{Grunhutetal2013}; \citealt{Wessolowskietal1988}; \citealt{LeuenhagenHamann1998}; further discussion is in section \ref{sec:summary}).
 \item The secondary can account for the amount of absorption in the  \ion{He}{1} lines, as we show in section \ref{sec:abs}.
 \item The Doppler shift of the emission follows that of the
absorption, as we show in section \ref{sec:orbital}.
\end{enumerate}

The secondary mass loss rate is much lower than the primary's. It is important to mention that most of the lines observed
from the $\eta$ Car system do originate in the primary and its wind (the best example is probably the hydrogen lines; \citealt{Weisetal2005}, \citealt{{Davidsonetal2005}}).
Take \ion{H}{1}~$\lambda4103\rm{\AA}$ line for example. This line, and many others, originate in the primary.
But it shows a P~Cyg profile that shifts much less in periastron compared to the \ion{He}{1} P~Cyg lines we discuss here.

It might seem a ``strange'' coincidence that the place in the secondary wind where the \ion{He}{1} are formed, according to our model, has about the
same velocity as the primary's wind.
However, even a stranger coincidence exists for a model where the lines are formed in the primary's wind: the area that absorbs the lines would have to 
change its velocity in the same way the secondary moves around the center of mass of the binary system.

Using the assumption that the \ion{He}{1} lines originate in the
secondary wind, in the past we fitted their Doppler shift
variations with orbital phase for the low-masses model (the conventional model) of $\eta$
Car, with $M_1=120 M_\odot$ and $M_2=30 M_\odot$
\citep{KashiSoker2008}. The Doppler shift of the P~Cyg absorption
component of the \ion{He}{1} lines was found to be in agreement
with the binary orientation with a longitude angle
$\omega=90^\circ$, i.e., secondary closest to us at periastron. We
here upgrade the model to include the new Doppler shifts presented
recently by \cite{Richardsonetal2015}, and the high-masses model
that better fits the luminosity and the behavior of $\eta$ Car
during the Great Eruption according to the HAPI model.

In fitting the \ion{He}{1} lines Doppler shifts with orbital phase we
scan the following parameter space.
\begin{enumerate}
\item{} The eccentricity is $e \simeq 0.85$--$0.93$.
\item{} Inclination angle (the angle between a line perpendicular to the orbital plane and the line of sight)
is also in the consensus to be $i \simeq 41^\circ$.
\item{} The orbital period is $P=2023$ days \citep{Daminelietal2008a}.
\item{} For the masses we use our results \citep{KashiSoker2010a} that the present masses are $M_1 \simeq 150$--$200 M_\odot$
and $M_2 \simeq 60$--$90 M_\odot$. We take here for the present masses of $\eta$ Car components $M_1=170 M_\odot$ and $M_2=80 M_\odot$.
\item{} We assume that the observer is behind the secondary at periastron, namely $\omega\simeq 90^\circ$.
\end{enumerate}

We note that there is inconsistency in some papers regarding the
exact epoch of periastron, or phase $0$. \cite{Richardsonetal2015}
used JD 2454842.5 for the 2009 periastron passage.
\cite{Mehneretal2011a} used JD 2454860 as a reference time for the event
(see also discussion at the appendix of \citealt{Mehneretal2011b}).
The uncertainty in determining the time of periastron may cause the change in radial velocity to
appear after periastron rather than before, or vice versa.
Evidently the observations collected so far from $\eta$ Car are insufficient for determining the exact time of periastron.
We will adopt an intermediate value between the references above, $t_{\rm{per}} =$ JD 2454850.
This value coincides with having the sharp variation in radial velocity of the lines studied here at periastron.
However, one should bear in mind that the uncertainty of the periastron epoch is of $\pm 10$ days.
We corrected the orbital phases inferred from the observation dates in this work to have phase 0 at $t_{\rm{per}}$.

\section{ORBITAL PARAMETERS}
\label{sec:orbital}

Based only on geometric considerations, i.e., neglecting
variations of the wind speed near periastron passage and
stochastic wind speed variations, we here use our model to fit the
observations of \cite{Richardsonetal2015}. The orbital velocity of
the secondary relative to the primary, $v_{\rm{orb}}$, is
converted to the velocity relative to the center of mass
\begin{equation}
  v_m = \frac{M_1}{(M_1+M_2)} v_{\rm{orb}} =
   \left[ \frac{G^{1/2} M_1}{(M_1+M_2)^{1/2}} \right]
   \left[ \frac{2}{r(t)}-\frac{1}{a} \right]^{1/2},
\label{v_m1}
\end{equation}
that can be written as
\begin{equation}
\begin{split}
  v_m &= \left[ \frac{M_1}{(M_1+M_2)^{2/3}} \right] \left[\frac{2 \pi G}{P}\right]^{1/3} \left[ \frac{2}{\tilde{r}(t)}-1 \right]^{1/2} \\
      &= f_M \left[\frac{2 \pi G}{P}\right]^{1/3} \left[ \frac{2}{\tilde{r}(t)}-1 \right]^{1/2}.
\end{split}
\label{v_m2}
\end{equation}
where $ \tilde{r}(t) \equiv r(t)/a$.

The entire stellar mass dependency is embedded in the factor
$f_M$. For the low-masses model ($M_1=120 M_\odot$ and $M_2=30
M_\odot$) its value is $f_M=4.25 M_\odot^{1/3}$, while for the
high-masses model ($M_1=170 M_\odot$ and $M_2=80 M_\odot$) it is
$f_M=4.28 M_\odot^{1/3}$. Therefore the amplitude of the fit with
the new parameters is less than $1\%$ larger than the amplitude
using the old parameters.

\begin{figure}[!t]
\centering
\includegraphics[trim= 0.0cm 0.0cm 0.0cm 0.0cm,clip=true,width=0.99\linewidth]{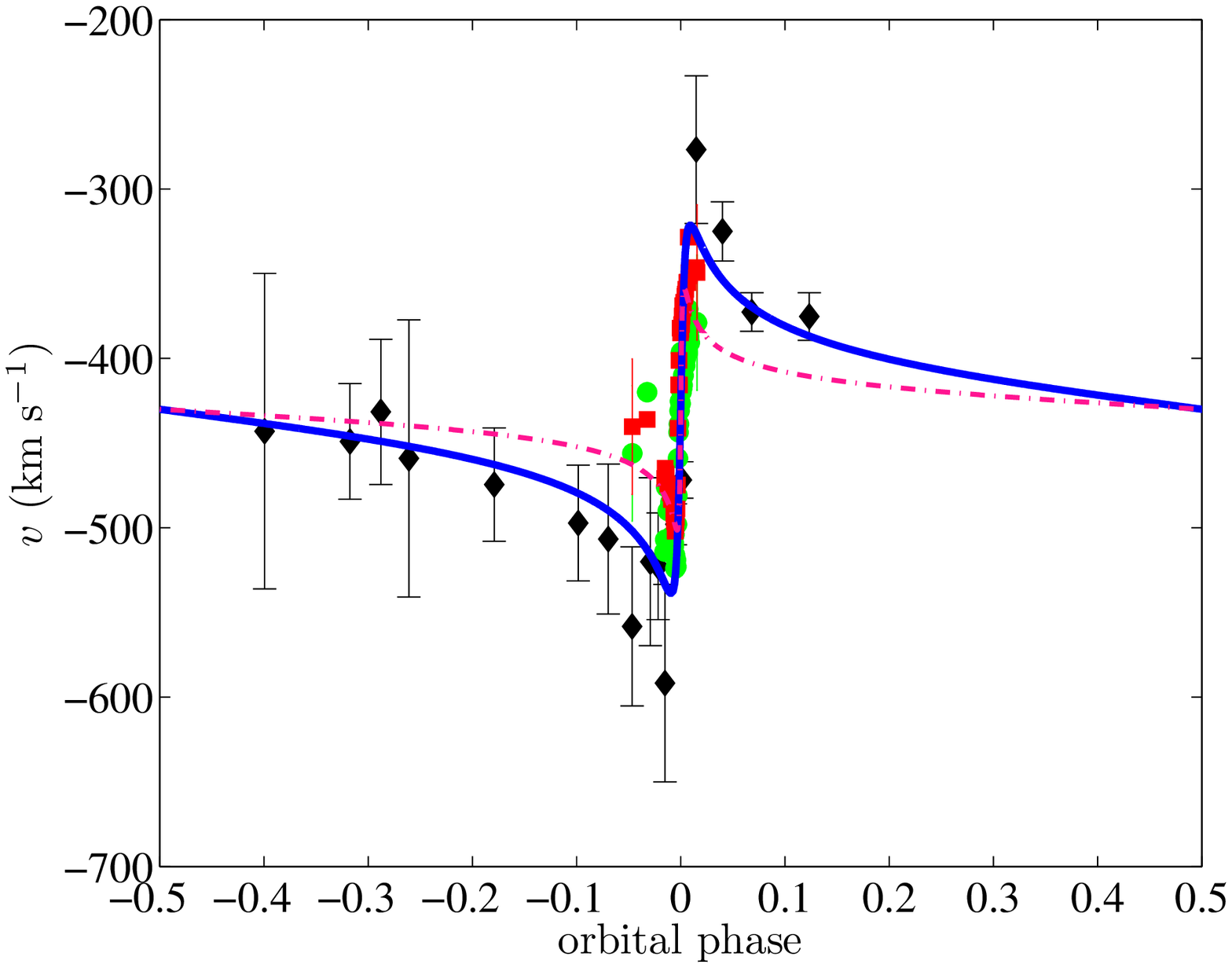}     
\includegraphics[trim= 0.0cm 0.0cm 0.0cm 0.0cm,clip=true,width=0.99\linewidth]{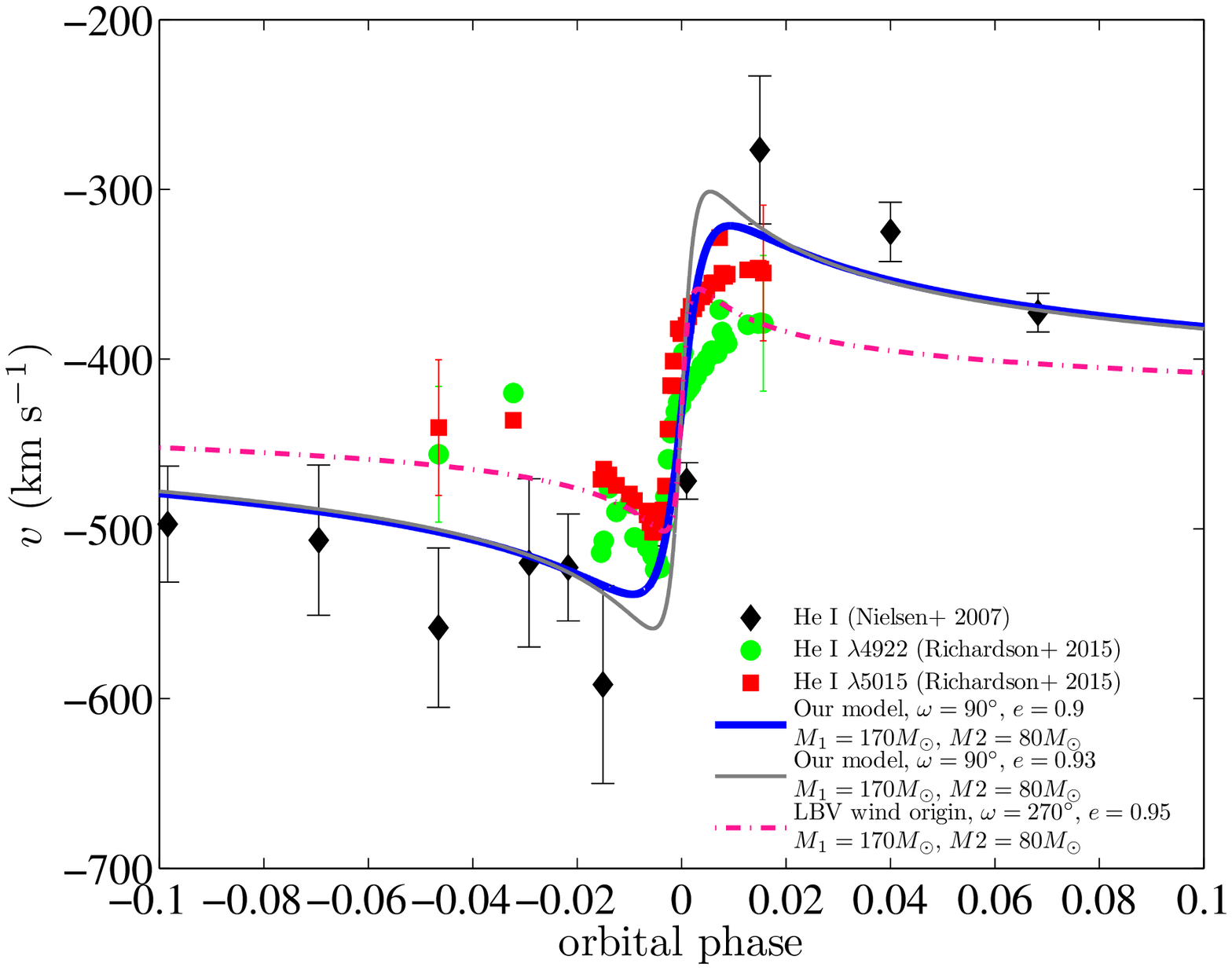}     
\caption{ Our fit to the observed radial velocity absorption
component of \ion{He}{1} lines. The black diamond points are the \ion{He}{1}~$\lambda7065\rm{\AA}$, $\lambda5876\rm{\AA}$,
$\lambda5015\rm{\AA}$, $\lambda4992\rm{\AA}$, and
$\lambda4471\rm{\AA}$ lines from
\cite{Nielsenetal2007} with the original error bars. The CTIO radial velocity data from
\cite{Richardsonetal2015} are shown in green
circles (\ion{He}{1}~$\lambda4922\rm{\AA}$) and red squares (\ion{He}{1}~$\lambda5015\rm{\AA}$),
and include no error bars. We added bars to two points of
each set that indicate our estimate of the fluctuations in the
velocity of the wind absorption component, $\pm 40 ~\kms$.
Observation phase was modified to periastron time defined in the
text. Our fiducial model assumes that the spectral lines originate
in the secondary stellar wind, that  $\omega = 90^\circ$ (i.e.,
secondary star closest to us at periastron), an eccentricity of
$e=0.9$, and stellar masses of $M_1=170 M_\odot$ and $M_2=80 M_\odot$.
The upper panel shows the entire
orbit, and the bottom panel is zoomed to times near periastron.
The lower panel includes also a model with $e=0.93$
instead of $e=0.9$ (grey line). For comparison we present a fit for a case
where the \ion{He}{1} lines originate from the primary stellar wind,
and the orientation is $\omega = 270^\circ$ (dashed-dotted pink line).
The velocity of the region in the wind responsible for the lines is taken to be
$v_{\rm{zone,abs}}=430~\kms$.}
 \label{fig:HeImin}
\end{figure}

On top of $v_m$ there are factors related to the observation angle
$\omega$, the inclination $i$, and the eccentricity $e$.  From the
result we subtract the constant velocity $v_{\rm{zone}}$ of the
zone in the secondary wind where the lines are absorbed, or the
average velocity of the emitting gas when the peak emission is
fitted.
Figure \ref{fig:HeImin} shows in a blue-solid line our fit to the
observed radial velocity absorption component of \ion{He}{1} lines
from \cite{Richardsonetal2015} and of \cite{Nielsenetal2007}.
It is clear from Figure \ref{fig:HeImin} that our fit to the new lines at
$\lambda4922\rm{\AA}$  and $\lambda5015\rm{\AA}$ is not as good as our fit
to the lines studied by \cite{Nielsenetal2007}. We attribute this to
contamination of the two $\lambda5015\rm{\AA}$ and $\lambda4922\rm{\AA}$
lines by \ion{Fe}{2} lines. This contamination was mentioned by
\cite{Richardsonetal2015}.
%

\begin{figure}[!t]
\centering
\includegraphics[trim= 0.0cm 0.0cm 0.0cm 0.0cm,clip=true,width=0.99\linewidth]{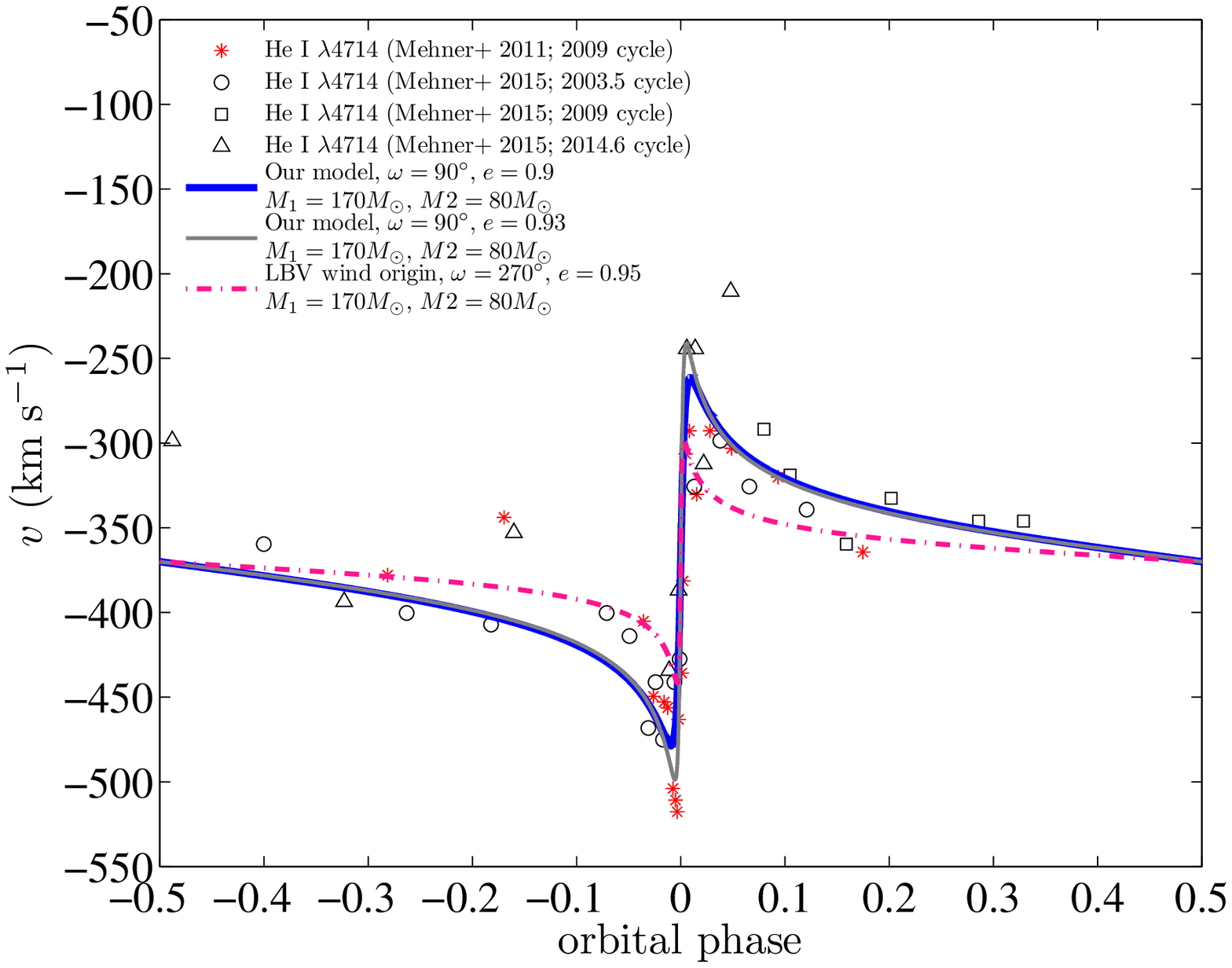}     
\caption{ Like Figure \ref{fig:HeImin}, but taking
$v_{\rm{zone,abs}}=370~\kms$ to fit the observed radial velocity
absorption component of the \ion{He}{1}~$\lambda4714\rm{\AA}$ line, from
observations of \cite{Mehneretal2011b} and \cite{Mehneretal2015}.
}
 \label{fig:HeI4714}
\end{figure}

More observations of \ion{He}{1} lines, specifically the
\ion{He}{1}~$\lambda4714\rm{\AA}$ line, were taken for the 2009
event by \cite{Mehneretal2011b}, and for the 2014.6 event by
\cite{Mehneretal2015}. The later paper summarizes observations of
that line from the previous 3 events (2003.5, 2009, and 2014.6).
Figure \ref{fig:HeI4714} shows how our model fits the
observations. We used the same model we used for the other
\ion{He}{1} lines in Figure \ref{fig:HeImin}, but
$v_{\rm{zone,abs}}=370~\kms$. We take this slightly different
value than the value of $v_{\rm{zone,abs}}=430~\kms$ that was used
in Figure \ref{fig:HeImin} to match the average of the absorption
component's radial velocity. The different values mean either
that the \ion{He}{1}~$\lambda4714\rm{\AA}$ line is absorbed in the
wind slightly closer to the wind origin on the secondary star, or
that there are large variations and uncertainties in the derived
Doppler shifts near periastron passages.

\cite{Mehneretal2015} found that the 2014.6 event was different
than pervious events, and according to their interpretation in the
framework of the accretion model, it showed signs of less
accretion onto the secondary close to periastron, indicating
weaker primary wind. The \ion{He}{1} line flux increased significantly in
2009--2014 compared to 1998--2003. The absorption of the
\ion{He}{1}~$\lambda4714\rm{\AA}$ line disappeared 8 days before
periastron and re-appeared 8 days after (assuming our
$t_{\rm{per}}$). The radial velocities are lower in the 2014.6
event compared to previous events. In fact, for fitting only the
data from 2014.6 it would be better to use
$v_{\rm{zone,abs}}=330~\kms$, keeping the rest of the parameters
unchanged. This behavior, together with the large fluctuations in
the Doppler shifts near periastron as seen in the different
figures, suggest that the velocities of the regions where lines
are formed in the different regions of the secondary wind vary
from cycle to cycle and in short time scales near periastron
passages. We should therefore aim at fitting the general behavior
of the Doppler shift variations with orbital phase; a perfect fit
to the Doppler shifts of this interacting binary system cannot be
achieved using the geometrical effects alone.

\begin{figure}[t!]
\centering
\includegraphics[trim= 0.0cm 0.0cm 0.0cm 0.0cm,clip=true,width=0.99\linewidth]{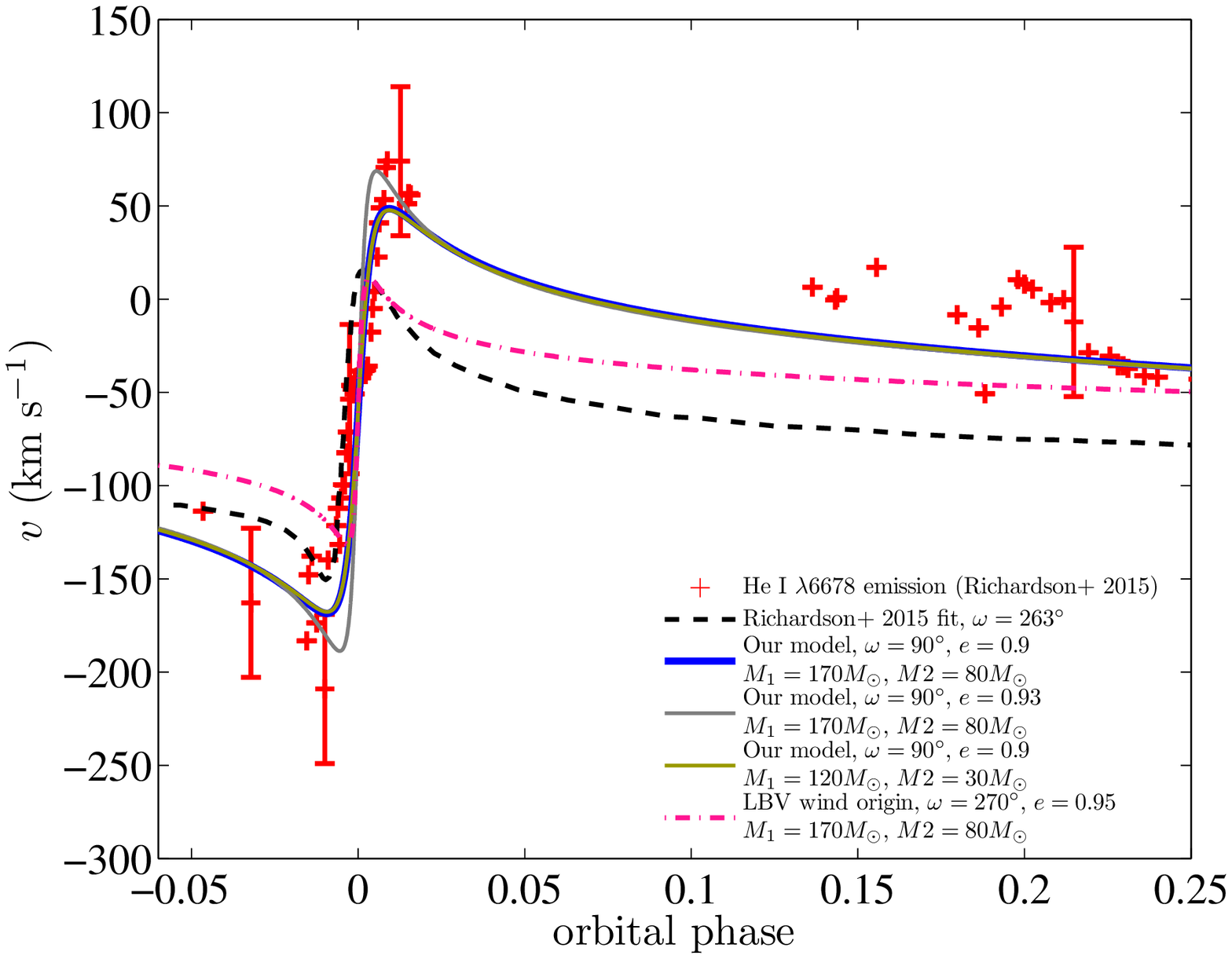}     
\caption{ Our fit to the emission peak bisector velocity of the
\ion{He}{1}~$\lambda6678\rm{\AA}$ line (blue line). We show also
the low-masses model (yellow line), and a higher eccentricity
model (grey line). Bars indicate our estimate of
fluctuations in the average velocity of the emitting gas in the
secondary wind. Also shown is the case where the line
originates in the primary stellar wind and the primary star is
closest to us at periastron (dashed-dotted pink line).
The black dashed line is the fit to the
data taken from \cite{Richardsonetal2015}, who assume that the
line originates in the primary wind.}
 \label{fig:HeIvb}
\end{figure}

We used the same principles we used for fitting the radial
velocity variations of the \ion{He}{1} absorption lines to fit the
\emph{emission} peak bisector velocity of the
\ion{He}{1}~$\lambda6678\rm{\AA}$ line across the 2009 event
observed by \cite{Richardsonetal2015}. We find that a value of
$v_{\rm{zone,emi}}=60 ~\kms$ for the average velocity of the gas
emitting the line  gives the best results. Figure \ref{fig:HeIvb}
shows the fit we obtained, together with the observations of the
emission peak bisector velocity of the
\ion{He}{1}~$\lambda6678\rm{\AA}$. We present two other cases with
$\omega=90^\circ$ (secondary star closer to the observer at periastron),
one with an eccentricity of $e=0.93$, and the other is the
low-masses model from \cite{KashiSoker2008}. We also show two
models where the line is assumed to originate in the primary
stellar wind. One is our simple geometric model, but the line is
emitted by the primary stellar wind and $\omega=270^\circ$
(dashed-dotted line), and the second is the fit presented by
\cite{Richardsonetal2015}. It is clear that the models where the
line is emitted by the secondary stellar wind and the secondary
star is in the foreground at periastron result in a much better
fit to the observations.
%

\begin{figure}[!t]
\centering
\includegraphics[trim= 0.0cm 0.0cm 0.0cm 0.0cm,clip=true,width=0.99\linewidth]{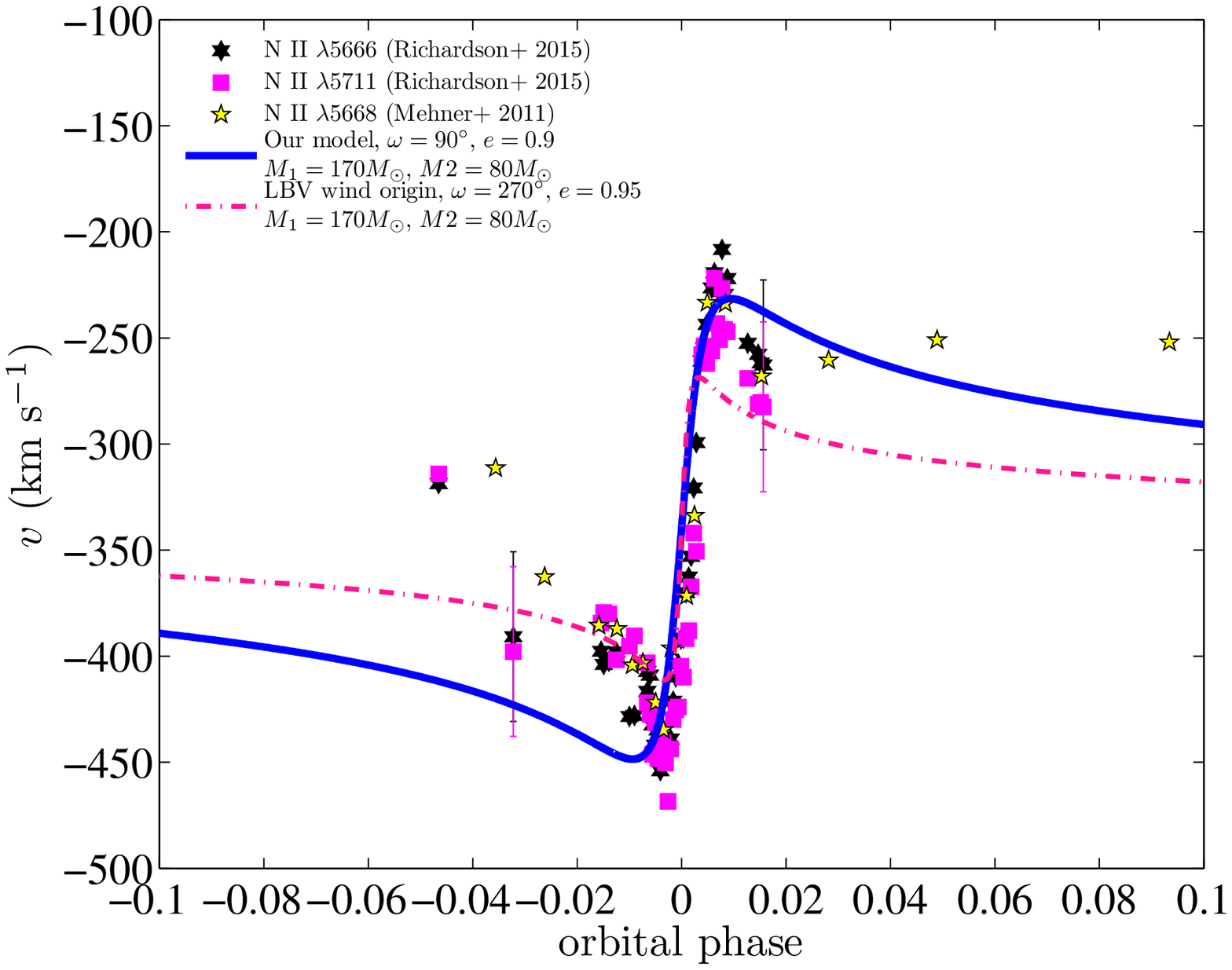}     
\includegraphics[trim= 0.0cm 0.0cm 0.0cm 0.0cm,clip=true,width=0.99\linewidth]{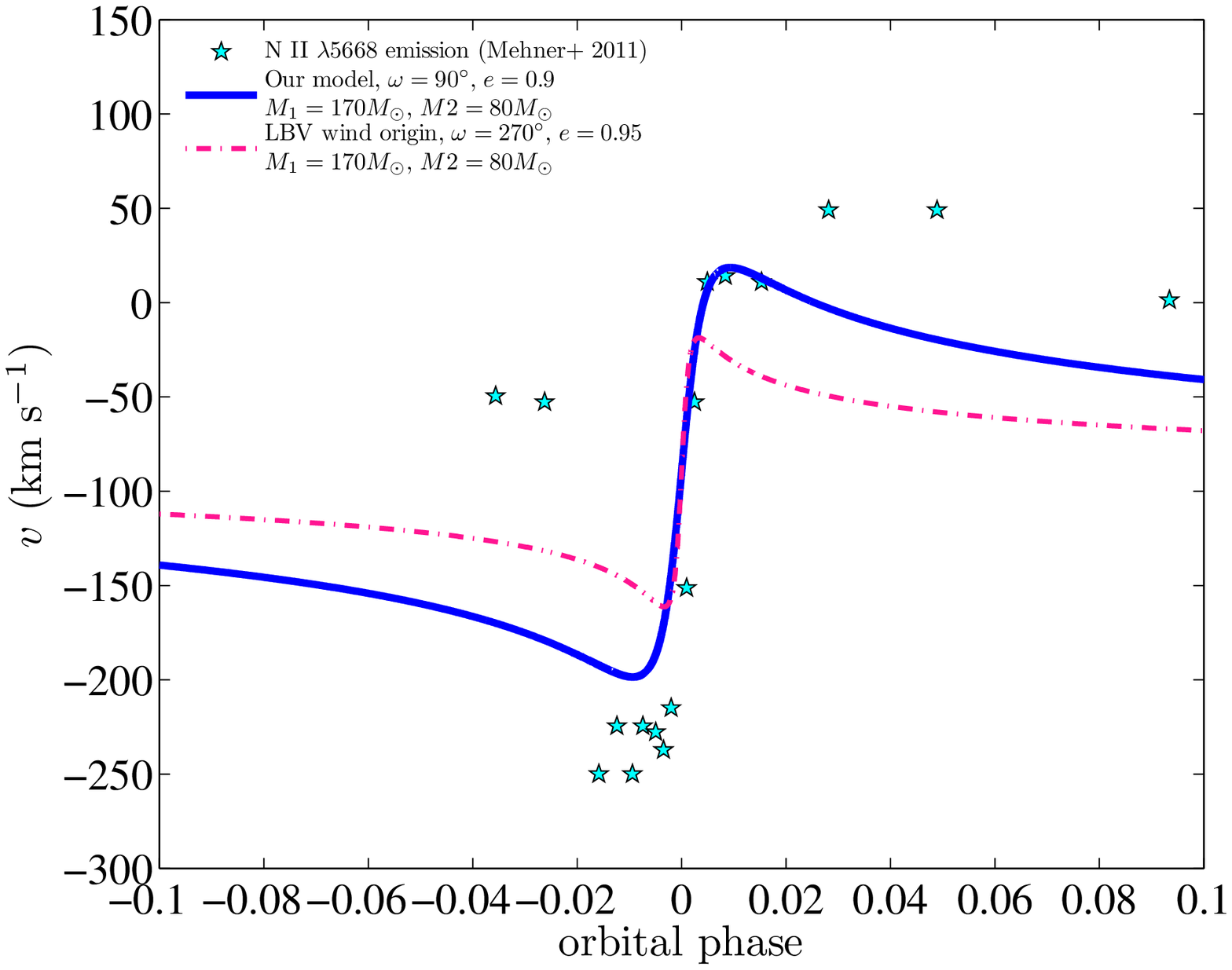}     
\caption{ Upper panel: Our fit to the observed radial velocity
absorption component of the \ion{N}{2}~$\lambda5666\rm{\AA}$ and \ion{N}{2}~$\lambda5771\rm{\AA}$ lines from the 2009 event
(\citealt{Richardsonetal2015}), and the \ion{N}{2}~$\lambda5668\rm{\AA}$
line observed by \cite{Mehneretal2011a} in the 3 previous events.
Lower panel: Our fit to the observed radial velocity emission
component of the \ion{N}{2}~$\lambda5668\rm{\AA}$ line from
\cite{Mehneretal2011a}. }
 \label{fig:NII}
\end{figure}

We next test our model against the variation in the Doppler shift
of the absorption component of the
\ion{N}{2}~$\lambda5666$--$5771\rm{\AA}$ lines, taken from both
\cite{Richardsonetal2015} and \cite{Mehneretal2011a}. We use here
$v_{\rm{zone,abs}}=370~\kms$.  The \ion{N}{2}~$\lambda5668\rm{\AA}$ line
observed by \cite{Mehneretal2011a} has clear emission peaks close
to periastron, but these were not clear in the profiles of
\cite{Richardsonetal2015}. We fit also the emission component of
the \ion{N}{2}~$\lambda5668\rm{\AA}$ line, with
$v_{\rm{zone,emi}}=70~\kms$. The fits are presented in Figure
\ref{fig:NII}.

Though the \ion{N}{2} and \ion{He}{1} lines intensities behave differently, the
radial velocity of the \ion{N}{2}~$\lambda5668\rm{\AA}$ emission
component follows that of the absorption component amazingly well.
Also, the entire lines-formation regions show the same Doppler
shift variation. Such a behavior cannot be explained in frame of a
model where the \ion{N}{2} and \ion{He}{1} lines-formation regions change their
location within the primary wind.

The main driver of the variations in the Doppler shifts is the
pure orbital motion. However, it is not the only one. A variation
in location within the secondary wind in our model can take place,
and it is even expected to occur. As the secondary approached
periastron, its wind properties change, most likely due to
accretion of gas from the primary stellar wind \citep{Soker2005,
Akashietal2013}. The accretion phase lasts for several weeks, but
the influence on the secondary stellar wind properties can last
few months after the event, until the accreted mass is removed by
the restoring secondary wind \citep{KashiSoker2009c}. It is
therefore very reasonable that lines will be absorbed in different
locations across periastron. This secondary effect is the main
cause for the deviation from our Doppler-based model, as seen in
the figures.

\section{ABSORPTION OF HELIUM BY THE SECONDARY}
\label{sec:abs}

According to \cite{Nielsenetal2007} the amount of absorption of the \ion{He}{1}~$\lambda5015\rm{\AA}$ P~Cyg line reaches up to $50\%$.
This line, however, shows different orbital change (probably because it is blended) and is \emph{not} one of the lines we claim to originate in the secondary wind (see section \ref{sec:orbital}).
The \ion{He}{1} lines we attribute to the secondary, show much less absorption.
The \ion{He}{1}~$\lambda7067\rm{\AA}$ P~Cyg line, for example, shows $10\%$ absorption near periastron \cite{Nielsenetal2007}.
We can use this to better constrain the properties of the two stars.

The maximum absorption, assuming it occurs in the secondary wind, is obtained when
\begin{equation}
A_{\rm{max}}=\frac{f_{l2}L_2}{f_{l_1}L1+f_{l2}L_2}
\label{eq:absorption}
\end{equation}
Where $L_1$ and $L_2$ are the luminosities of the primary and the secondary, respectively,
and $f_{l_1}$ and $f_{l_1}$ are their fractions within the line absorption waveband.

According to \cite{Nielsenetal2007}, the \ion{He}{1}~$\lambda7067\rm{\AA}$ shows $10\%$ absorption for a range between $-700$--$-500~\kms$.
Taking $A_{\rm{max}}\leq0.1$ in equation (\ref{eq:absorption}) we get the requirement that 
\begin{equation}
\frac{f_{l1}}{f_{l2}} \frac{L_1}{L_2} \leq 9 .
\label{eq:absorption2}
\end{equation}
Assuming Blackbody radiation, and taking conventional parameters for the temperatures of the primary and the secondary $T_1=20\,000 \rm{K}$ and $T_2=40\,000 \rm{K}$, respectively,
we obtain $f_{l1}/f_{l2}=5.9$.
This means that according to equation (\ref{eq:absorption2}), the luminosity ratio needs to satisfy $L_2/L_1 \geq 0.65$.
According to the conventional parameters, the ratio $L_2/L_1=0.2$, so this leads to a contradiction.
It is not possible to find for the conventional model, a point in the stellar evolution path, even with a different effective temperature,
that would be even close to satisfy that requirement.
We therefore conclude that the two stars must have a smaller luminosity ratio, and consequently smaller mass ratio, as is expected in the massive stars model.
They should also have a smaller temperature ratio to make the requirement in equation (\ref{eq:absorption2}) easier to meet.

For the massive stars model we propose, we can track the evolution of $M_1=170 M_\odot$ and $M_2=80 M_\odot$ stars (e.g., \citealt{Ekstrometal2012}, \citealt{Kohleretal2015} and references therein), and
try to find a reasonable set of parameters that satisfies the above equation.
We find that if we take $T_1=25\,000 \rm{K}$ \citep{Hillieretal2001}, $T_2=37\,000 \rm{K}$ \citep{Verneretal2005}, which gives $f_{l1}/f_{l2}=2.8$,
then the requirement from equation (\ref{eq:absorption2}) becomes $L_2/L_1 \geq 0.31$.
And indeed, stellar evolution tracks give $L_1\simeq 3\times 10^6 L_\odot$ and $L_2\simeq 1.2\times 10^6 L_\odot$, satisfying the condition.

\section{SUMMARY AND DISCUSSION}
\label{sec:summary}

We used spectroscopic observations of $\eta$ Car close to the 2009
periastron passage (\citealt{Richardsonetal2015}; \citealt{Mehneretal2011a},\citeyear{Mehneretal2011b})
to show they support earlier suggestions that companion is in the
foreground at periastron (\citealt{KashiSoker2011} and references therein).

We assumed that the \ion{He}{1} and \ion{N}{2} spectral lines
originate in the acceleration zone of the secondary star of $\eta$
Car (\citealt{KashiSoker2007}, \citeyear{KashiSoker2011}).
We then took the secondary to be closest to us at periastron passages, i.e., $\omega=90^\circ$. We
further used the high-masses model of $\eta$ Car with component
masses of $M_1=170 M_\odot$ and $M_2=80 M_\odot$. These masses
better fit evolutionary tracks of massive stars that cross the
locations of the two stars on the HR diagram, than the commonly
used masses of $M_1=120 M_\odot$ and $M_2=30 M_\odot$.
A massive primary star with initial mass larger than $M_1=200 M_\odot$,
and a secondary star with an initial mass larger than $M_2=50 M_\odot$ are 
supported by stellar evolution calculations for very massive stars
(e.g., \citealt{Yungelsonetal2008}; \citealt{Brottetal2011}; \citealt{Yusofetal2013}).

The high masses, more generally in the range of $M_1 \simeq 150$--$200 M_\odot$
and $M_2 \simeq 60$--$90 M_\odot$, can account
also for the powering of the nineteenth century
Great Eruption by mass accretion onto the secondary star (the HAPI
model; \citealt{KashiSoker2010a}). For the eccentricity and
inclination angle we used the commonly accepted values of $e
\simeq 0.9$ and $i=41^\circ$, respectively. As evident from Figs.
\ref{fig:HeImin}--\ref{fig:NII} we could fit the general variation
of the Doppler shifts with orbital phase. Therefore, the
suggestion that the secondary star is in the foreground at
periastron (e.g., \citealt{KashiSoker2008, Tsebrenkoetal2013}) is
definitely tenable. The suggestion that the primary star is in
the foreground at periastron seems to encounter problems,
e.g., as evident from the dashed black line in Figure
\ref{fig:HeI4714} that is the model proposed by
\cite{Richardsonetal2015}.

In some cases the opposite model, of a line originating from the
primary stellar wind and the primary star is in the foreground
at periastron passages, might fit part of the Doppler shifts,
e.g., fitting the data from \cite{Richardsonetal2015} in Figure
\ref{fig:HeImin}. However, this model has too low amplitudes both
at periastron passages, as evident from all figures, and away from
periastron passages, e.g., right side of Figure \ref{fig:HeI4714}.

In the present study we considered only the role of the orbital
motion on the variation of the Doppler shift with orbital phase.
It appears clear from the fluctuations in the Doppler shift values and
from the non-perfect fitting that the velocity of the zone
responsible for the formation of each line is changing. Both
stochastic variations and variation with orbital phase, noticeably
near periastron passages, exist. These variations are another
manifestation of the unrelaxed nature of this binary system.

The three unknowns about the binary system discussed here are:
(i) masses of two stars; (ii) orbital orientation; (iii)
exact periastron time.
The masses of the two stars were obtained by \cite{KashiSoker2010a}
as explained in section \ref{sec:intro}.
Here we showed that the mass estimate obtained by \cite{KashiSoker2010a} also allows
to fit radial velocities of the lines.
The orbital orientation is the main parameter discussed here in detail.
The time of periastron is uncertain. We used a fixed value here
and did not fine-tuned its value.
It may however be possible to use the He I and other lines to
get a better constrain for its time. What we find here is that
our fits are consistent with periastron time of JD 2454850
but a few days difference is also possible.

In addition to the Doppler shifts presented here, there are other
arguments that support the suggestion that the secondary star is
in the foreground at periastron passages. The four supporting
arguments listed by \cite{KashiSoker2008} include the evolution of
the radio emission and the behavior of the
\ion{He}{1}~$\lambda10830\rm{\AA}$ line. In a previous study
\citep{KashiSoker2009b} we further argued that the column density
toward the X-ray emitting gas, that is the post-shock secondary
wind, is more compatible with a binary orientation where for most
of the time the secondary star is in the background, being on the
foreground only near periastron passages. Another supporting
argument was brought by \cite{Tsebrenkoetal2013}. They
demonstrated that the asymmetric morphology of the blue and
red-shifted components of the outflow at hundreds of astronomical
units from $\eta$ Car, can be accounted for from the collision
of the free primary stellar wind with the slowly expanding dense
equatorial gas closer to us. Namely, for most of the orbital
period the primary is in the foreground, and at periastron
passages the secondary star is in the foreground.

\cite{Humphreysetal2008} discovered that the \ion{He}{1} lines were absent from $\eta$ Car's spectra
prior to the mid 1940s.
They discussed the various difficulties it may pose to models attributing the required He ionizing photons to the
the secondary, ionizing the primary wind.
They also conclude that even a much denser primary wind could not have obscured the He ionizing photons coming from the secondary.
\cite{Clementeletal2015b} and \cite{Mehneretal2015} claimed, however, that the He ionizing photons only moderately penetrate the dense
post-shocked primary wind. Therefore if the primary wind at that time was 2--4 times denser, it should have been enough to change the ionization structure
of He in the primary's wind, and prevent the formation of the lines.
These arguments, however, are irrelevant for a model in which the \ion{He}{1} lines originate in the secondary's wind.
In the frame of the accretion model, however, it is easier to provide an explanation.
A denser primary wind can form a thick accretion belt around the secondary close to periastron \citep{KashiSoker2009c} that
would last for the entire orbit, providing a shield for its radiation.
Even after the belt is gone, the mass that is accreted onto the secondary changes its photospheric structure and makes it cooler,
diminishing the He ionizing photons.
This may require that the primary wind before the 1940s was 20--30 times denser.
A new study by \cite{Kashietal2016} suggests that the mass loss could have reached that magnitude or even higher.

\cite{Mehneretal2011a} also observed the \ion{N}{2}
lines and their velocity shifts from reflected polar spectra at the location known as ``FOS4''.
They argue it may be a problematic observation for the orbital motion explanation to the Doppler shift variations.
A similar argument appear in \cite{Mehneretal2011b} regarding the \ion{He}{1}~$\lambda4714\rm{\AA}$ line.
The only direct comparison of FOS4 and direct view is in figure 8 of that paper.
Though it may at first sight look like the radial velocity of the absorption in FOS4 follows that of the direct view,
we notice the following:
(1) The observations at $-353$ and $-82$ days (phase $-0.18$ and $-0.045$) show very high radial velocity in absorption, much above other observations of the same line.
(2) At $-82$ days the value of FOS4 is $40~\km \s^{-1}$ higher. As the observations are sparse, there is now way to know if this is significant.
It may indicate that the FOS4 Doppler shift of the absorption is smaller in amplitude and non systematic. Alternatively, it may be a fluctuation.
It is therefore impossible to know if FOS4 persistently follows the direct view or not.
A detailed comparison of densely sampled multiple lines is needed in order to check that.
Even if it does, it may well be possible that not only polar light is reflected to FOS4 and there is some reflection from equatorial regions.

\cite{Mehner2010} used the distribution of gas and ionizing radiation around $\eta$ Car to
constrain the properties of the secondary.
If the limits of \cite{Mehner2010} hold, then the secondary mass should be $M_2 \lesssim 60 M_\odot$.
A mass of $M_2=60 M_\odot$ is still within the HAPI model for the Great Eruption.
We here showed that the Doppler shifts can be fitted with $M_2=30 M_\odot$ and $M_2=80 M_\odot$
Any value for $M_2$ in this range can be fitted. But as stated,
because of the luminosity of the primary star and the HAPI model for the GE,
we prefer the high-masses model.

P Cyg profiles in He~I are found in hot hydrogen poor stars 
(\citealt{Leuenhagenetal1996}; \citealt{Wessolowskietal1988}; \citealt{LeuenhagenHamann1998}).
We know from observations that the primary's outer layers consists of about $50\%$ helium (\citealt{Davidsonetal1986}; \citealt{Dufouretal1997}).
A few $M_\odot$ of material from the primary were accreted onto the secondary during the eruptions.
The accreted gas makes the secondary's envelope enriched with helium.
It is very plausible that even though the secondary is hot, its helium lines are stronger than other stars in its evolutionary stage.
Clumpiness of the secondary wind \citep{KashiSoker2007} can make part of the gas somewhat cooler, also enhancing the He~I lines.

The very massive primary star of $M_1>150 M_\odot$ and the very high eccentricity of
the binary orbit hints that the system was once a triple system,
and that the primary formed by the merger of two (or more) stars.
The merger process released large amount of gravitational energy
within weeks to months. Such an event can be classified as ILOT.
Therefore, it may well be that the nineteenth century Great Eruption
was not the first ILOT of this system.
We note that the estimates for the masses in the HAPI model do not depend on the previous existence or nonexistence of a third star.
The masses of only two stars are relevant for both the calculations done in this paper for spectral fitting, and the calculation
done in \cite{KashiSoker2010a} for modeling the light curve.
The third star is suggested as a possible easier route to obtain the large mass of the primary, together with a high eccentricity orbit.
But it is well possible that both were obtained with only two stars along the entire evolution.

\vspace*{0.3cm}
AK acknowledges support provided by National Science Foundation through grant AST-1109394.
We thank Roberta Humphreys,  Kris Davidson, and an anonymous referee for helpful comments.


\vspace*{0.1cm}

\begin{thebibliography}{}

\bibitem[Abraham \& Falceta-Gon{\c c}alves(2007)]{AbrahamFalceta2007} Abraham, Z., \& Falceta-Gon{\c c}alves, D.\ 2007, \mnras, 378, 309

\bibitem[Abraham et al.(2014)]{Abrahametal2014} Abraham, Z., Falceta-Gon{\c c}alves, D., \& Beaklini, P.~P.~B.\ 2014, \apj, 791, 95

\bibitem[Abraham et al.(2005)]{Abrahametal2005} Abraham, Z., Falceta-Gon{\c c}alves, D., Dominici, T.~P., Nyman L.-A. D. P., McAuliffe F., Caproni A., Jatenco-Pereira V.\ 2005, \aap, 437, 977

\bibitem[Akashi et al.(2013)]{Akashietal2013} Akashi, M.~S., Kashi, A., \& Soker, N.\ 2013, \na, 18, 23

\bibitem[Akashi et al.(2006)]{Akashietal2006} Akashi, M., Soker, N., \& Behar, E.\ 2006, \apj, 644, 451

\bibitem[Brott et al.(2011)]{Brottetal2011} Brott, I., de Mink, S.~E., Cantiello, M., et al.\ 2011, \aap, 530, A115 

\bibitem[Clementel et al.(2015a)]{Clementeletal2015a} Clementel, N., Madura, T.~I., Kruip, C.~J.~H., \& Paardekooper, J.-P.\ 2015a, \mnras, 450, 1388

\bibitem[Clementel et al.(2015b)]{Clementeletal2015b} Clementel, N., Madura, T.~I., Kruip, C.~J.~H., Paardekooper, J.-P., \& Gull, T.~R.\ 2015b, \mnras, 447, 2445

\bibitem[Corcoran(2005)]{Corcoran2005} Corcoran, M.~F.\ 2005, \aj, 129, 2018

\bibitem[Corcoran et al.(2010)]{Corcoranetal2010} Corcoran, M.~F., Hamaguchi, K., Pittard, J.~M., Russell, C.~M.~P., Owocki, S.~P., Parkin, E.~R., \& Okazaki, A.\ 2010, \apj, 725, 1528

\bibitem[Chen et al.(2015)]{Chenetal2015} Chen, Y., Bressan, A., Girardi, L., et al.\ 2015, \mnras, 452, 1068

\bibitem[Crowther \& Bohannan(1997)]{CrowtherBohannan1997} Crowther, P.~A., \& Bohannan, B.\ 1997, \aap, 317, 532

\bibitem[Damineli(1996)]{Damineli1996} Damineli, A.\ 1996, \apjl, 460, L49

\bibitem[Damineli et al.(1997)]{Dminelietal1997} Damineli, A., Conti, P.~S., \& Lopes, D.~F.\ 1997, \na, 2, 107

\bibitem[Damineli et al.(2008a)]{Daminelietal2008a} Damineli, A., Hillier, D.~J., Corcoran, M.~F., et al.\ 2008a, \mnras, 384, 1649

\bibitem[Damineli et al.(2008b)]{Daminelietal2008b} Damineli, A., Hillier, D.~J., Corcoran, M.~F., Stahl O., Groh J.~H., Arias J., Teodoro M., \& Morrell N.\ 2008b, \mnras, 386, 2330

\bibitem[Davidson et al.(1986)]{Davidsonetal1986} Davidson, K., Dufour,  R.~J., Walborn, N.~R., \& Gull, T.~R.\ 1986, \apj, 305, 867

\bibitem[Davidson et al.(2005)]{Davidsonetal2005} Davidson, K., Martin, J., Humphreys, R.~M., et al.\ 2005, \aj, 129, 900

\bibitem[Davidson \& Humphreys(1997)]{DavidsonHumphreys1997} Davidson, K., \& Humphreys, R.~M.\ 1997, \araa, 35, 1

\bibitem[Davidson et al.(2015)]{Davidsonetal2015} Davidson, K., Mehner, A., Humphreys, R.~M., Martin, J.~C., \& Ishibashi, K.\ 2015, \apjl, 801, L15

\bibitem[Dufour et al.(1997)]{Dufouretal1997} Dufour, R.~J., Glover,  T.~W., Hester, J.~J., et al.\ 1997, Luminous Blue Variables: Massive Stars  in Transition, 120, 255

\bibitem[Duncan \& White(2003)]{DuncanWhite2003} Duncan, R.~A., \& White, S.~M.\ 2003, \mnras, 338, 425

\bibitem[Ekstr{\"o}m et al.(2012)]{Ekstrometal2012} Ekstr{\"o}m, S., Georgy, C., Eggenberger, P., et al.\ 2012, \aap, 537, A146

\bibitem[Falceta-Gon{\c c}alves et al.(2007)]{Falcetaetal2007} Falceta-Gon{\c c}alves, D., Abraham, Z., \& Jatenco-Pereira, V.\ 2007, IAU Symposium, 240, 198

\bibitem[Falceta-Gon{\c c}alves et al.(2005)]{Falcetaetal2005} Falceta-Gon{\c c}alves, D., Jatenco-Pereira, V., \& Abraham, Z.\ 2005, \mnras, 357, 895

\bibitem[Figer et al.(1998)]{Figeretal1998} Figer, D.~F., Najarro, F., Morris, M., McLean, I.~S., Geballe, T.~R., Ghez, A.~M., \& Langer, N.\ 1998, \apj, 506, 384

\bibitem[Georgy et al.(2012)]{Georgyetal2012} Georgy, C., Ekstr{\"o}m, S., Meynet, G., et al.\ 2012, \aap, 542, A29

\bibitem[Gomez et al.(2010)]{Gomezetal2010} Gomez, H.~L., Vlahakis, C., Stretch, C.~M., Dunne, L., Eales, S.~A., Beelen, A., Gomez, E. L., \& Edmunds, M.~G.\ 2010, \mnras, 401, L48

\bibitem[Groh et al.(2012)]{Grohetal2012} Groh, J.~H., Hillier, D.~J., Madura, T.~I., \& Weigelt, G.\ 2012, \mnras, 423, 1623

\bibitem[Groh et al.(2010)]{Grohetal2010} Groh, J.~H., Nielsen, K.~E., Damineli, A., et al.\ 2010, \aap, 517, A9

\bibitem[Grunhut et al.(2013)]{Grunhutetal2013} Grunhut, J.~H., Wade, G.~A., Leutenegger, M., et al.\ 2013, \mnras, 428, 1686 

\bibitem[Gull et al.(2011)]{Gulletal2011} Gull, T.~R., Madura, T.~I., Groh, J.~H., \& Corcoran, M.~F.\ 2011, \apjl, 743, L3

\bibitem[Hamaguchi et al.(2007)]{Hamaguchietal2007} Hamaguchi, K., Corcoran, M.~F., Gull, T., et al.\ 2007, \apj, 663, 522

\bibitem[Hamaguchi et al.(2014a)]{Hamaguchietal2014a} Hamaguchi, K., Corcoran, M.~F., Russell, C.~M.~P., et al.\ 2014a, \apj, 784, 125

\bibitem[Hamaguchi et al.(2014b)]{Hamaguchietal2014b} Hamaguchi, K., Corcoran, M.~F., Takahashi, H., et al.\ 2014b, \apj, 795, 119

\bibitem[Henley et al.(2008)]{Henleyetal2008} Henley, D.~B., Corcoran, M.~F., Pittard, J.~M., et al.\ 2008, \apj, 680, 705

\bibitem[Hillier et al.(2001)]{Hillieretal2001} Hillier, D.~J., Davidson, K., Ishibashi, K., \& Gull, T.\ 2001, \apj, 553, 837

\bibitem[Humphreys et al.(2008)]{Humphreysetal2008} Humphreys, R.~M., Davidson, K., \& Koppelman, M.\ 2008, \aj, 135, 1249

\bibitem[Iping et al.(2005)]{Ipingetal2005} Iping, R.~C., Sonneborn, G., Gull, T.~R., Massa, D.~L., \& Hillier, D.~J.\ 2005, \apjl, 633, L37

\bibitem[Kashi et al.(2016)]{Kashietal2016} Kashi, A., Davidson, K., \& Humphreys, R.~M.\ 2016, \apj, 817, 66 

\bibitem[Kashi \& Soker(2007)]{KashiSoker2007} Kashi, A., \& Soker, N.\ 2007, \na, 12, 590   

\bibitem[Kashi \& Soker(2008)]{KashiSoker2008} Kashi, A., \& Soker, N.\ 2008, \mnras, 390, 1751

\bibitem[Kashi \& Soker(2009a)]{KashiSoker2009a} Kashi, A., \& Soker, N.\ 2009a, \apjl, 701, L59

\bibitem[Kashi \& Soker(2009b)]{KashiSoker2009b} Kashi, A., \& Soker, N.\ 2009b, \mnras, 397, 1426

\bibitem[Kashi \& Soker(2009)]{KashiSoker2009c} Kashi, A., \& Soker, N.\ 2009, \na, 14, 11

\bibitem[Kashi \& Soker(2010a)]{KashiSoker2010a} Kashi, A., \& Soker, N.\ 2010a, \apj, 723, 602

\bibitem[Kashi \& Soker(2010b)]{KashiSoker2010b} Kashi, A., \& Soker, N.\ 2010b, arXiv:1011.1222

\bibitem[Kashi \& Soker(2011)]{KashiSoker2011} Kashi, A., \& Soker, N.\ 2011, arXiv:1104.4655

\bibitem[Kashi \& Soker(2015)]{KashiSoker2015} Kashi, A., \& Soker, N.\ 2015, RAA, (arXiv:1508.00004)

\bibitem[K{\"o}hler et al.(2015)]{Kohleretal2015} K{\"o}hler, K., Langer, N., de Koter, A., et al.\ 2015, \aap, 573, A71

\bibitem[Leuenhagen et al.(1996)]{Leuenhagenetal1996} Leuenhagen, U., Hamann, W.-R., \& Jeffery, C.~S.\ 1996, \aap, 312, 167

\bibitem[Leuenhagen \& Hamann(1998)]{LeuenhagenHamann1998} Leuenhagen, U., \& Hamann, W.-R.\ 1998, \aap, 330, 265

\bibitem[Livio \& Pringle(1998)]{LivioPringle1990} Livio, M., \& Pringle, J.~E.\ 1998, \mnras, 295, L59 

\bibitem[Madura et al.(2015)]{Maduraetal2015} Madura, T.~I., Clementel, N., Gull, T.~R., Kruip, C.~J.~H., \& Paardekooper, J.-P.\ 2015, \mnras, 449, 3780

\bibitem[Madura et al.(2013)]{Maduraetal2013} Madura, T.~I., Gull, T.~R., Okazaki, A.~T., Russell, C.~M.~P., Owocki, S.~P., Groh, J.~H., Corcoran, M.~F., Hamaguchi, K., \& Teodoro, M.\ 2013, \mnras, 436, 3820

\bibitem[Madura et al.(2012)]{Maduraetal2012} Madura, T.~I., Gull, T.~R., Owocki, S.~P.,  Groh, J.~H., Okazaki, A.~T., Russell, C.~M.~P.\ 2012, \mnras, 420, 2064

\bibitem[Matt \& Balick(2004)]{MattBalick2004} Matt, S., \& Balick, B.\ 2004, \apj, 615, 921 

\bibitem[Mehner et al.(2010)]{Mehner2010} Mehner, A., Davidson, K., Ferland, G.~J., \& Humphreys, R.~M.\ 2010, \apj, 710, 729

\bibitem[Mehner et al.(2011a)]{Mehneretal2011a} Mehner, A., Davidson, K., \& Ferland, G.~J.\ 2011, \apj, 737, 70

\bibitem[Mehner et al.(2011b)]{Mehneretal2011b} Mehner, A., Davidson, K., Martin, J.~C., et al.\ 2011, \apj, 740, 80

\bibitem[Mehner et al.(2010)]{Mehneretal2010} Mehner, A., Davidson, K., Ferland, G. J., Humphreys, R. M. 2010, ApJ, 710, 729

\bibitem[Martin et al.(2010)]{Martinetal2010} Martin, J.~C., Davidson, K., Humphreys, R.~M., \& Mehner, A.\ 2010, \aj, 139, 2056

\bibitem[Mehner et al.(2012)]{Mehneretal2012} Mehner, A., Davidson, K., Humphreys, R.~M.,  et al.\ 2012, \apj, 751, 73

\bibitem[Mehner et al.(2015)]{Mehneretal2015} Mehner, A., Davidson, K., Humphreys, R.~M., et al.\ 2015, \aap, 578, A122

\bibitem[Nielsen et al.(2007)]{Nielsenetal2007} Nielsen, K.~E., Corcoran, M.~F., Gull, T.~R., Hillier, D. J., Hamaguchi, K., Ivarsson, S., \& Lindler, D.~J.\ 2007, \apj, 660, 669

\bibitem[Okazaki et al.(2008)]{Okazakietal2008} Okazaki, A.~T., Owocki, S.~P., Russell, C.~M.~P., \& Corcoran, M.~F.\ 2008, \mnras, 388, L39

\bibitem[Parkin et al.(2009)]{Parkinetal2009} Parkin, E.~R., Pittard, J.~M., Corcoran, M.~F., Hamaguchi, K., \& Stevens, I.~R.\ 2009, \mnras, 394, 1758

\bibitem[Pittard \& Corcoran(2002)]{PittardCorcoran2002} Pittard, J.~M., \& Corcoran, M.~F.\ 2002, \aap, 383, 636

\bibitem[Pittard et al.(1998)]{Pittardetal1998} Pittard, J.~M., Stevens, I.~R., Corcoran, M.~F., \& Ishibashi, K.\ 1998, \mnras, 299, L5

\bibitem[Portegies Zwart \& van den Heuvel(2016)]{PortegiesZwartvandenHeuvel2016} Portegies Zwart, S.~F., \& van den Heuvel, E.~P.~J.\ 2016, \mnras, 456, 3401 

\bibitem[Richardson et al.(2015)]{Richardsonetal2015} Richardson, N.~D., Gies, D.~R., Gull, T.~R., Moffat, A.~F.~J., \& St-Jean, L.\ 2015, arXiv:1507.07417

\bibitem[Smith(2006)]{Smith2006} Smith, N.\ 2006, \apj, 644, 1151

\bibitem[Smith(2010)]{Smith2010} Smith, N.\ 2010, \mnras, 402, 145

\bibitem[Smith \& Ferland(2007)]{SmithFerland2007} Smith, N., \& Ferland, G.~J.\ 2007, \apj, 655, 911

\bibitem[Smith \& Frew(2011)]{SmithFrew2011} Smith, N., \& Frew, D.~J.\ 2011, \mnras, 415, 2009

\bibitem[Smith et al.(2000)]{Smithetal2000} Smith, N., Morse, J.~A., Davidson, K., \& Humphreys, R.~M.\ 2000, \aj, 120, 920

\bibitem[Soker(2005)]{Soker2005} Soker, N.\ 2005, \apj, 635, 540

\bibitem[Soker(2001)]{Soker2001} Soker, N.\ 2001, \mnras, 325, 584

\bibitem[Teodoro et al.(2016)]{Teodoroetal2016} Teodoro, M., Damineli, A., Heathcote, B., et al.\ 2016, \apj, 819, 131 

\bibitem[Tsebrenko et al.(2013)]{Tsebrenkoetal2013} Tsebrenko, D., Akashi, M., \& Soker, N.\ 2013, \mnras, 429, 294

\bibitem[Verner et al.(2005)]{Verneretal2005} Verner, E., Bruhweiler, F., \& Gull, T.\ 2005, \apj, 624, 973 

\bibitem[Weis et al.(2005)]{Weisetal2005} Weis, K., Stahl, O., Bomans, D.~J., et al.\ 2005, \aj, 129, 1694 

\bibitem[Wessolowski et al.(1988)]{Wessolowskietal1988} Wessolowski, U., Schmutz, W., \& Hamann, W.-R.\ 1988, \aap, 194, 160

\bibitem[Whitelock et al.(2004)]{Whitelocketal2004} Whitelock, P.~A., Feast, M.~W., Marang, F., \& Breedt, E.\ 2004, \mnras, 352, 447

\bibitem[Yungelson et al.(2008)]{Yungelsonetal2008} Yungelson, L.~R., van den Heuvel, E.~P.~J., Vink, J.~S., Portegies Zwart, S.~F., \& de Koter, A.\ 2008, \aap, 477, 223

\bibitem[Yusof et al.(2013)]{Yusofetal2013} Yusof, N., Hirschi, R., Meynet, G., et al.\ 2013, \mnras, 433, 1114



\end{thebibliography}
\end{document}